 \documentclass[preprint2]{aastex}

\shorttitle{The Detection of a Light Echo from SN 2007af}
\shortauthors{Drozdov et al.}

\pdfoutput=1 
\usepackage{natbib}
\usepackage{hyperref}
\usepackage{float}
\usepackage[caption=false]{subfig}
\usepackage{graphicx}
\usepackage{pdfpages}
\usepackage{upgreek}
\usepackage{threeparttable}
\usepackage{amsmath}
\usepackage{float}
\usepackage{epstopdf}
\usepackage{footnote}
\usepackage[bottom]{footmisc}
\usepackage{epsf} 
\begin{document}

\title{Detection of a Light Echo from the Otherwise Normal SN 2007af}

\author{D. Drozdov\altaffilmark{1}, M. D. Leising\altaffilmark{1}, P. A. Milne\altaffilmark{2}, J. Pearcy\altaffilmark{2}, A. G. Riess\altaffilmark{3}, L. M. Macri\altaffilmark{4}, G. L. Bryngelson\altaffilmark{5}, P. M.  Garnavich\altaffilmark{6}}

\altaffiltext{1}{Department of Physics and Astronomy, Clemson University, Clemson, SC 29634, USA.}
\altaffiltext{2}{University of Arizona, Steward Observatory, Tucson, AZ 85719, USA.}
\altaffiltext{3}{Department of Physics and Astronomy, Johns Hopkins University, Baltimore, MD 21218, USA}
\altaffiltext{4}{George P. and Cynthia Woods Mitchell Institute for Fundamental Physics and Astronomy, Department of Physics \& Astronomy, Texas A\&M University, 4242 TAMU, College Station, TX 77843, USA}
\altaffiltext{5}{Department of Physics and Astronomy, Francis Marion University, Florence, SC 29501, USA}
\altaffiltext{6}{Physics Department, University of Notre Dame, Notre Dame, IN 45556, USA}


\begin{abstract}
We present the discovery of a light echo from SN 2007af, a normal Type Ia supernova (SN Ia) in NGC 5584. \textit{Hubble Space Telescope} ($HST$) images taken three years post explosion reveal two separate echoes; an outer echo and extended central region, which we propose as an unresolved inner echo.  Multiple images were obtained in the F160W, F350LP, F555W, and F814W using the Wide Field Camera 3.  If the outer echo is produced by an interstellar dust sheet perpendicular to the line of sight, it is located $\sim$800 pc in front of the SN. The dust for the inner echo is 0.45 pc $\textless$ d $\textless$ 90 pc away from the SN. The inner echo color is consistent with typical interstellar dust wavelength-dependent scattering cross-sections, while the outer echo color does not match the predictions.  Both dust sheets, if in the foreground, are optically thin for scattering, with the outer echo sheet thickness consistent with the inferred extinction from peak brightness. Whether the inner echo is from interstellar or circumstellar dust is ambiguous.  Overall, the echo characteristics are quite similar to previously observed SN Ia echoes. 
\end{abstract}




\section{INTRODUCTION}

Type Ia supernovae have long been invoked as ``standard candle" distance indicators due to their extreme luminosity and relative homogeneity of that luminosity. Although powerful as tools in astronomy, there are fundamental properties of the underlying physics that are not yet fully understood. Single-degenerate (SD) and double-degenerate (DD) progenitor binary systems have been proposed to produce SNe Ia and various explosion models have been suggested to describe these seemingly uniform objects.  Among the approaches to discern new information about supernovae and their environments, light echoes can be effective probes of both the circumstellar and interstellar environments. 

Light echoes (LEs) are rare events, where the light from a bright source  scatters off surrounding dust and reaches the observer after a delay caused by the extra distance traveled. Evidence of light echoes were first discovered from Nova Persei 1901 \citep{Ritchey_a,Ritchey_b,Ritchey02}.  Since that time, light echoes have been seen from Galactic Nova Sagittarii  \citep{Swope},  Galactic Cepheid RS Puppis  \citep{Havlen}, V838 Monocerotis \citep{Bond}, young stars \citep{Ortiz, Soko} and other phenomena. Some examples from Type II supernovae include the most famous case of SN 1987A \citep{Crotts, Gou, Sun, Panagia}, SN 1993J  \citep{Liu03}, SN 1999ev  \citep{Van,1999ev}, SN 2003gd  \citep{2003gd}, SN 1980K  \citep{Sugerman12}, SN 2004et \citep{Otsuka}, SN 2006gy \citep{Miller}, and SN 2008bk \citep{VanDyk}.

Light echoes can be used to determine various dust properties of the host galaxy \citep{Patat05}. Time-variability of spectral features near peak brightness suggests the presence of circumstellar material (CSM). The CSM could be due to stellar winds and/or non-conservative mass transfer from the donor star. The presence of CSM has often been treated as evidence of a single degenerate progenitor system, but recent studies suggest that both SD and DD scenarios can produce CSM \citep{Moore, Raskin13, Shen}. The mass loss history and the progenitor system can be investigated from the detection of a CSM light echo. 

The SN-dust distance can be found simply from the geometry of an interstellar echo.   Because they reveal the dusty nature of the surroundings, light echoes can also be used to probe the suggested association of some SNe Ia with star-forming regions, and therefore, the young nature of the progenitors \citep{Howell}.  The distance to the host galaxy can also be determined geometrically from the ring of maximum linear polarization from echoes \citep{Sparks94}.  In practice, this method is limited by the point spread function (PSF) of the image.

Light echoes have also been used to classify historical SNe \citep{Rest08, Krause_b}. Light echoes are dominated by scattered peak light, and the light-curve weighted mean of the spectrum  can be compared to standard SNe spectra for classification purposes.  Understanding the prevalence and characteristics of light echoes is essential for late-time monitoring of SN Ia power since weak echoes might be misinterpreted (e.g., as dissipation of trapped positron kinetic energy, when the light curve falls ten or more magnitudes below peak brightness).

Four Type Ia supernova light echoes have been reported to date (SN 1991T \citep{Schmidt_1991T}, SN 1998bu \citep{Capp,Garn}, SN 1995E \citep{QuinnGarn}, and SN 2006X \citep{Wang08, Crotts6X}). Here we report the detection of a light echo from SN 2007af discovered in $HST$ images and corroborated with ground-based observations. The paper is organized as follows.  Early- and late-epoch observations of SN 2007af are discussed in Section~\ref{sec:sn2007af}. The light echo analyses and interpretations are located in Sections~\ref{sec:LE} and~\ref{sec:phys}, respectively. Section~\ref{sec:disc} focuses on the interstellar versus circumstellar dust argument for the inner echo production, and we present our conclusions in Section~\ref{sec:conc}.

\section{SN 2007AF}
\label{sec:sn2007af}

SN 2007af was discovered by K. Itagaki on 2007 March 1.84 (UT dates are used throughout) at $\alpha = 14h 22m 21.03s$, and $\delta = -0^{\circ} 23' 37.6''$ well before maximum.  The location of the SN was offset 40$''$ west and 22$''$ south from the dusty center of the host galaxy, NGC 5584. NGC 5584 is an Scd galaxy with a recession velocity of 1638 km $s^{-1}$ \citep{Koribalski}. SN 2007af was classified as a normal SN Ia by its spectrum taken on 2007 March 4.34 \citep{Salgado}.   

 \citet{Simon} reported on the lack of time-variability of the Na D absorption features in the spectra of SN 2007af, in contrast to the variability observed from SN 2006X by \citet{Patat07}. High-resolution spectra were taken at -4.3, +16.6, and +23.7 days relative to maximum light using the ARCES echelle spectrograph on the ARC 3.5m telescope at Apache Point Observatory. Na D absorption lines and the H$\alpha$ emission line were extensively investigated, but no variability was detected. Due to the lack of variation in the spectral absorption features, the paper concludes that absorbing gas from the interstellar medium of NGC 5584 is responsible for the Na D absorption lines. The progenitor system either differs from the SD circumstellar case of SN 2006X or spectroscopic variability is not detectable along all viewing directions. The lack of time-variability constrains the location of the circumstellar material, but does not rule out its existence.

The line-of-sight extinction estimate from \citet{Simon} was A$_{V}$ = 0.39 $\pm$ 0.06 mag using an extinction law of R$_{V}$ = 2.98 $\pm$ 0.33. Due to the low extinction and the normal behavior at peak, SN 2007af would be considered an unlikely candidate for a light echo. 

SN 2007af was the target of multiple photometric campaigns due to its discovery at an early epoch. Ultraviolet, near infrared, and optical photometry were performed on this normal SN Ia by a variety of teams and telescopes. The separation of the SN from the center of the host galaxy and the relatively nearby galaxy provided an ideal SN for intensive study.

\subsection{STEWARD OBSERVATIONS}

We monitored SN 2007af with the Montreal 4K imager (Mont4K) on the 1.5m Kuiper telescope and the 90Prime Imager on the 2.3m Bok telescope.  The 61$''$ Kuiper telescope is located on Mount Bigelow in the Santa Catalina Mountains.  The Mont4K optical imager is equipped with a Fairchild 4000 x 4000 CCD with a 9$'$.7 x 9$'$.7 field of view. The telescope features a primary focal ratio of $f$/13.5 Cassegrain focus.  The optical imager used on the Bok telescope on Kitt Peak, 90Prime, is a prime focus, wide-field imager, which utilizes a mosaic array of four 4000 x 4000 pixel CCDs and images an area of 1.0 square degrees. Aperture photometry was performed using Landolt standard stars \citep{Landolt}, and the images were reduced using standard IRAF\footnote[1]{The Image Reduction and Analysis Facility (IRAF) is publicly distributed by the National Optical Astronomy Observatory (NOAO) in Tucson, AZ. NOAO is operated by the Association of Universities for Research in Astronomy, Inc. in cooperation with the National Science Foundation. } procedures. The late-epoch Steward observations were analyzed after template subtraction using both \textit{HST} images and images obtained from NASA/IPAC Extragalactic Database (NED)\footnote[2]{The NASA/IPAC Extragalactic Database (NED) is operated by the Jet Propulsion Laboratory, California Institute of Technology, under contract with the National Aeronautics and Space Administration.} . 

\begin{table*}[hp]
\centering
\caption{\hspace{90pt}Optical Photometry from Steward Observatory}
\label{table:stew}
\begin{tabular}{ccccccccccccc}
\hline \hline
\\JD (days) & B & err(B) & V & err(V)& R & err(R) \\
\hline \hline \\
2454208.9 & 16.199 & 0.004 & 14.985 & 0.001 & 14.543 &	0.002 \\
2454229.5 & 16.709 & 0.004 & 15.694 & 0.002 & 15.352 & 0.002\\
2454230.5 & 16.702 & 0.004 & 15.737 & 0.003 & 15.443 & 0.002 \\
2454231.5 & 16.792 & 0.004 & 15.734 & 0.002 & 15.461 & 0.002 \\
2454246.5 & 16.878 & 0.004 & 16.059 & 0.002 & 15.872 & 0.004 &\\
2454257.5 & 17.030 & 0.007 & 16.573 & 0.007 & 16.200 & 0.003 \\
2454481.5 & 20.390 & 0.048 & 20.381 & 0.040 & 20.422 & 0.407 \\
2454509.5 & ... & ...	& 20.250 & 0.040 & 20.888 & 0.070\\
2454539.5 & ... & ...	& 20.830 & 0.065 & 21.315 & 0.081 \\
2454540.5 & ... & ...	& 20.780 & 0.030 & ... & ... 	\\
2454562.5 & 21.253 & 0.068	 & 21.537 & 0.076	& 21.514 & 0.080\\
2454908.5 & ... & ... & 22.762 & 0.462 & ... &	\\
\hline \hline
\end{tabular}
  \end{table*}

\subsection{$\textit{\textbf{HST}}$ OBSERVATIONS}

To use high-redshift SNe Ia in cosmology, the absolute magnitudes of SNe Ia must be calibrated.  The best calibrations of SNe Ia use Cepheid variable stars in the host galaxies of low-redshift SNe Ia, which are calibrated with parallax and main sequence fitting, two nearby rungs of the distance ladder.  Cepheid variables have also produced the most accurate measurement of $H_{o}$, the present expansion rate of the universe.  The ``Supernovae and $H_{o}$ for the Equation of State" (SHOES) project aimed to measure the expansion rate to $\textless$ 5\% precision using SNe Ia and Cepheids. SNe Ia chosen for this cosmological study had to fit the following criteria:  have modern photometric data (e.g. CCD), have observations before peak, low reddening ($A_{V}$ \textless \hspace{1pt} 0.5 mag), spectroscopically classified as normal, and have \textit{HST} optical observations of Cepheid variables in the SN host galaxy \citep{Riess}.      

NGC 5584 was discovered to have a wealth of Cepheids and was observed with the WFC3 during \textit{HST} Cycle 17 as part of $HST$ program 11570: P.I. A. Riess.  NGC 5584 was observed in the F160W (wide $H$), F350LP (unfiltered long pass), F555W (wide $V$), and F814W (wide $I$) filters from January -- April 2010, three years after the explosion of the SN. The individual exposures were 400 -- 700s in length with a total exposure time equaling 4926s. Integer and half-pixel dithering were enabled between exposures to characterize the PSF. 

\subsection{LIGHT CURVE}

\begin{figure}[htp]
\begin{center}
\includegraphics[width=3.in]{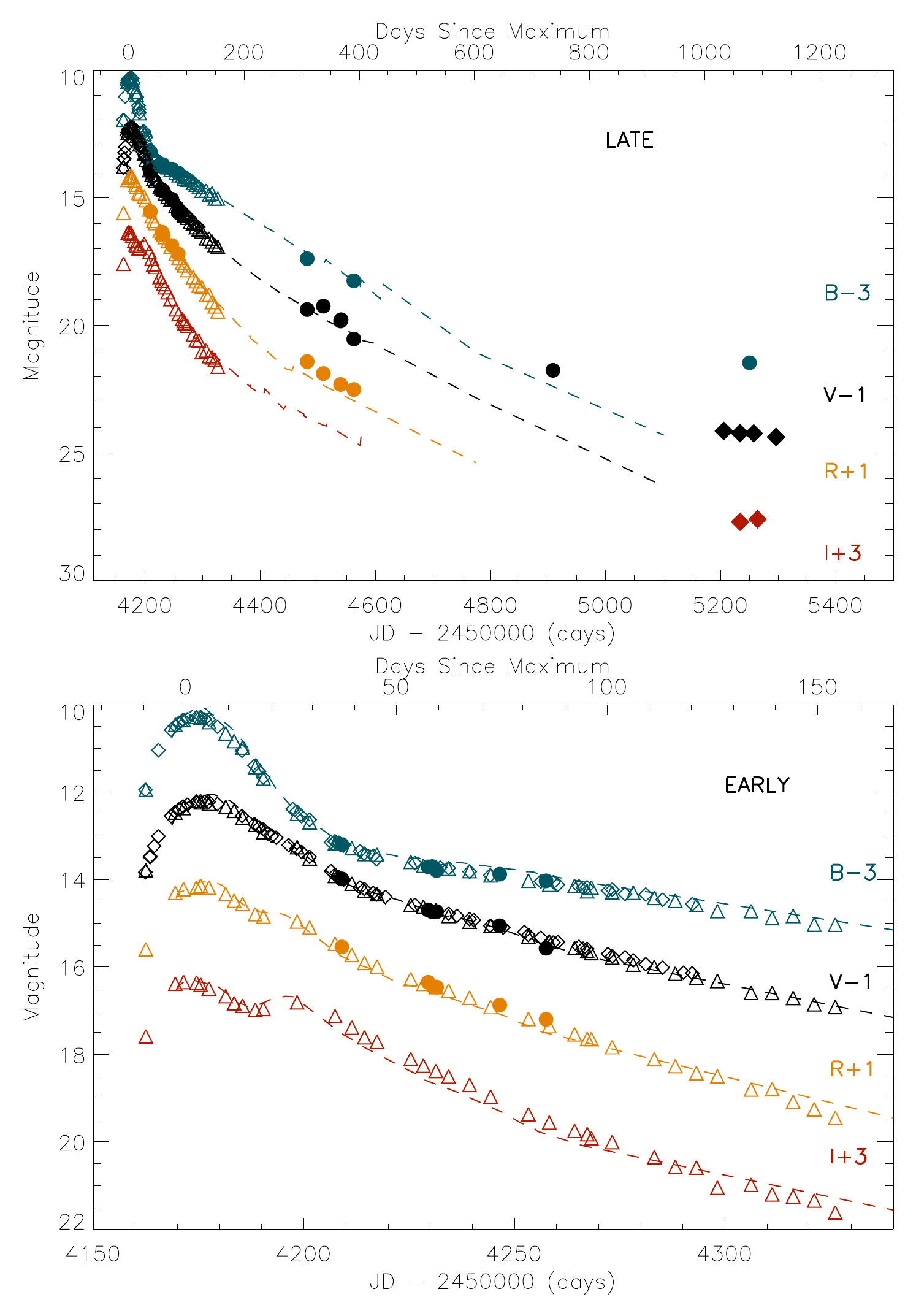}
\end{center}
\caption[SN 2007af Light Curve]{Early (lower panel)- and complete (upper panel) $BVRI$ light curves of SN 2007af plotted against normal SN 1992A (dashed line)\protect\citep{Sunt}. The light echo dominates at late-times. The \textit{HST} images are labeled with black diamonds (F555W) and red diamonds (F814W).  Steward observations are distinguished with filled circles.  The early data was taken from CfA3 ($BV$-open diamonds) and KAIT ($BVRI$-open triangles) \protect\citep{Hicken,Simon}.}
\label{fig:LC}
\end{figure}

Early- and late-epoch observations complete the light curve (Figure~\ref{fig:LC}) and are plotted against the dashed line of normal Type Ia SN 1992A \citep{Sunt}. At early epochs, the light curve shows no deviation from the normal template. The normal peak behavior of SN 2007af is in sharp contrast to the late plateau.  The 2009 March 18.2 (JD = 2454908.5) observations, two years past explosion, show the SN is considerably brighter than SN 1992A at a comparable epoch. The emission from SN 1992A during that epoch is thought to be due to the energy deposition from positrons created in $^{56}$Co $\rightarrow$ $^{56}$Fe decays \citep{Sunt, Mi}. The obvious flattening of the SN 2007af light curve compared to the decline rate from intrinsic emission is attributed to the existence of a light echo. \citet{Patat05} has argued that light echoes should be 10 -- 12 magnitudes fainter than the SN at peak, which is consistent with the observational sample.  A light echo 10 magnitudes fainter than peak will reveal itself at 450+ days post-maximum. 

The distinct flattening is present in $V$ and $I$, suggesting that an echo was captured in both the $HST$ images and ground-based Steward Observatory late-epoch observations in $V$.  SN 2007af was well observed in the transition from intrinsic to light echo emission.  At over 1000 days after peak, the SN is $\sim$4 magnitudes brighter than predicted from intrinsic emission.          

\section{LIGHT ECHO IMAGING} 
\label{sec:LE}

\begin{figure}[h]
\begin{center}
\includegraphics[width=2.5in]{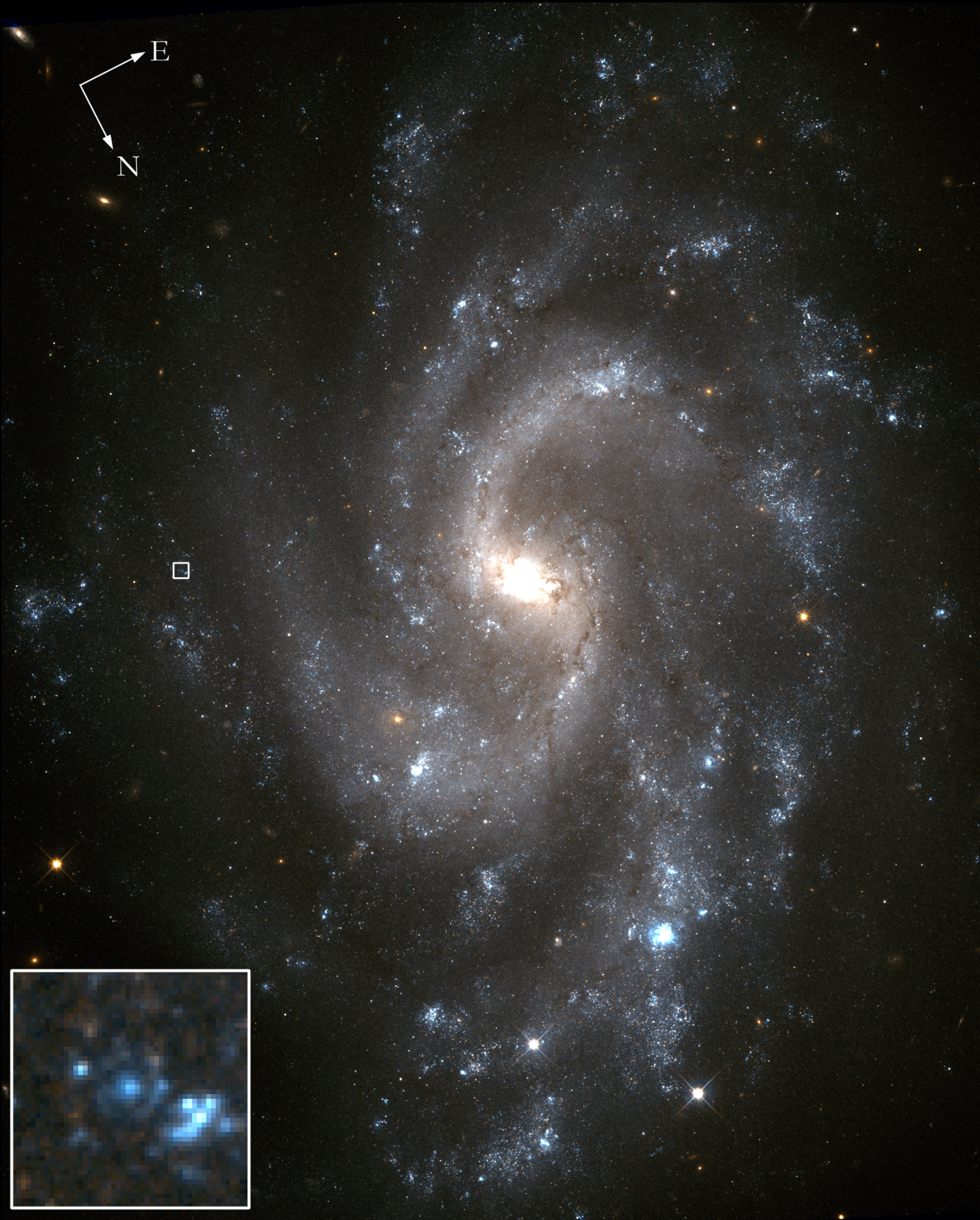}
\end{center}
\caption[NGC5584 and SN 2007af Light Echo]{Color composite image of NGC 5584 and the combined light echo image of SN 2007af (inset panel) from $HST$ WFC3.  The SN location (box) is shown in NGC 5584, and the 2$''$ x 2$''$ close up of the region shows both echoes. The images are the result of stacking the individual monthly observations in all filters. A ring (outer echo) surrounds the bright central region, which we propose as a secondary light echo (inner echo).}
\label{fig:evo350}
\end{figure}

The \textit{HST} images reveal the light echo visually (Figure~\ref{fig:evo350}).  These  $\sim$1000 days post explosion images show a ring-like object and extended central source at the location of the SN, which we call outer and inner light echoes. 

\begin{figure}[h]
\begin{center}
\includegraphics[width=2.0in]{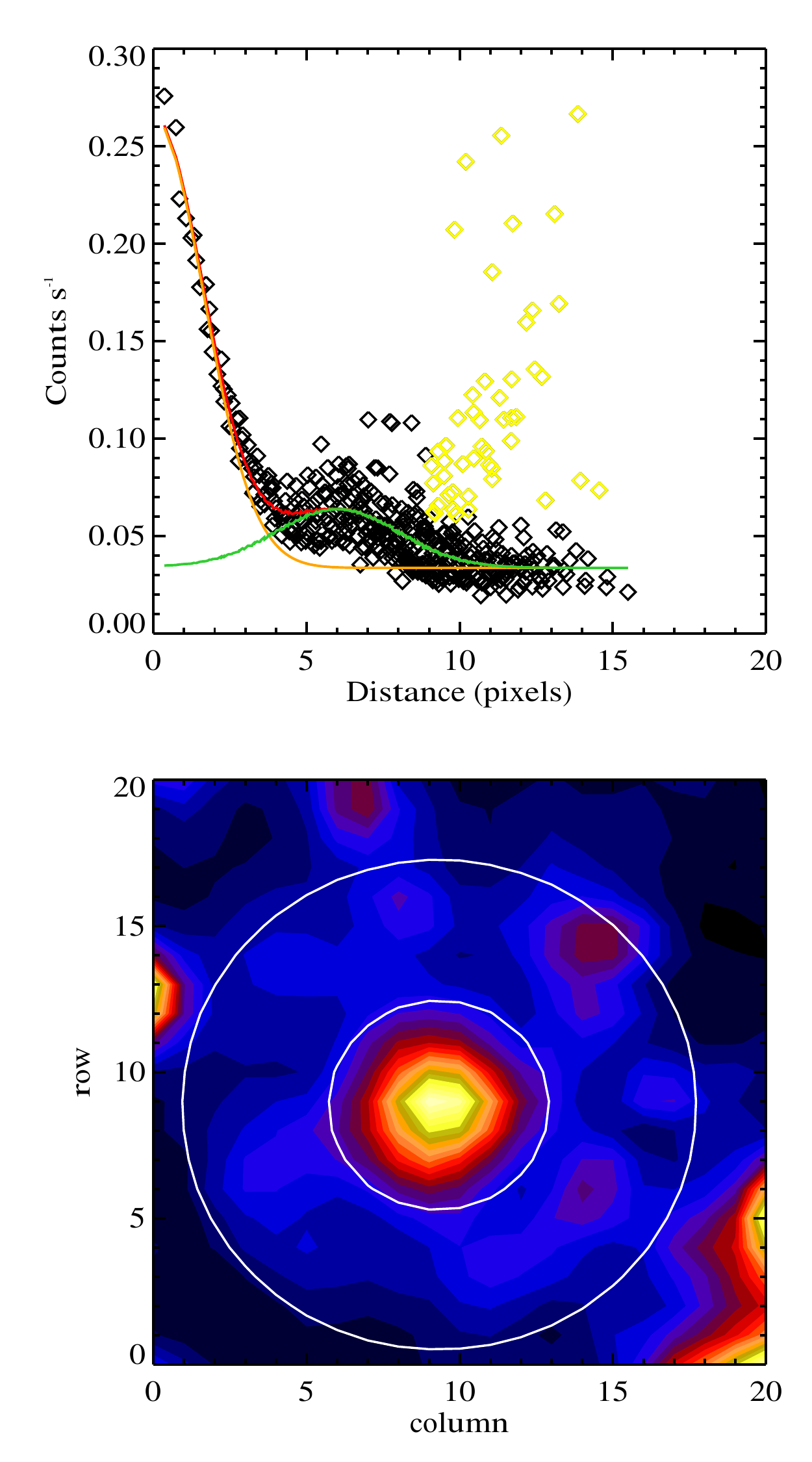}
\end{center}
\caption[]{An example of the fit of our echoes in the F350LP filter.  The top panel shows the count contributions in our crowded field.  The yellow diamonds are high points outside the outer echo that have been subtracted, the green line is the inner echo contribution, the orange line is the outer echo contribution, and red is inner and outer echo combined.  The bottom panel shows both echoes, where the white circles are 50$\%$ contours of the outer echo. The axes for both panels are pixel numbers from an arbitrary point on the CCD.}
\label{fig:fit}
\end{figure}

The Cepheid campaign began in January 2010, 2.8 years after the SN peak brightness. Thirty-nine images total were stacked to improve the signal-to-noise ratio in four filters, with the light echoes detected in all but the F160W($IR$) filter.  The echoes are clearly weaker in the F814W filter than in F350LP and F555W.  The echoes are faint, and in particular, the surface brightness of the outer feature is very low.  The region contains a number of other emission features, both stellar and otherwise. We characterized the echoes by fitting defined shapes convolved with the PSF's of the images (see Figure~\ref{fig:fit}).  

In this analysis, we select a limited region around the SN location, determine the PSF size from a number of stars (typically twenty), subtract stellar objects near the SN, and perform non-linear least-squares fits of two shapes  plus background to a region typically twenty-five pixels square around the SN. The PSF's are assumed to be azimuthally symmetric radial gaussians and are found to be consistent with  0.$''$11 full-width half-maximum (FWHM). We tried a variety of different shapes for the echo features. Fits of elliptical functions did not yield significant eccentricity, regardless of axis orientations, so we quote only fits for azimuthally symmetric functions. For the central source, we tried a flat plateau and exponentials, but no shape fit significantly better than a two dimensional radial gaussian, so we use that. For the outer annular echo, we use a radial function that falls off as a gaussian from a radius of maximum.

Our general fit has ten parameters: a constant background, central gaussian amplitude and width, gaussian annulus amplitude, peak radius, and width, and two coordinates of the center of each of the echo shapes. Convergence is unpredictable with all parameters free because of correlations among them, so we generally constrain some of them. If we try to determine the locations of the centroids of the echoes by fixing the widths of the two echoes, we always find the coordinates of both echo centers are consistent with the location of the SN and with each other, strengthening the argument of two separate echo detections. The locations are within one-half pixel of the SN position with a one-sigma uncertainty of typically 0.2 pixels. To determine the intensities and widths of the features, we constrain the coordinates of the centers of the two to be the same for both. Also, the two widths and the annulus radius are highly correlated for our quoted sizes and fluxes, so we fix the annulus FWHM at the mean best-fit value (0.$''$16) and allow the central source FWHM and annulus peak radius to vary.

Where there are non-stellar features in the fit region that are unrelated to the SN echoes, we exclude those points from the fits. The models are convolved with the PSF's and fit to the data. The resulting least-squares fits are typically not formally good fits, with reduced $\chi^2$ typically 1.5 to 2.0 for 450 degrees of freedom. The central echo is generally well fit, but the outer echo is not smooth and not very well fit by a smooth azimuthally symmetric echo. Still, the fits capture well the sizes and brightnesses of the echo components.  The angular radii measurements for the outer ($\theta_{OE}$) and inner echoes ($\theta_{IE}$) can be found in Table~\ref{table:width} for each filter,

\begin{table*}[htbp]
 \caption[Angular Sizes of Echoes]{\hspace{144pt}Angular Sizes of Echoes}
\label{table:width}
\centering
\begin{tabular}{cccccc}
\hline \hline
\\Filter & JD (days) & $\theta_{OE}('')$ & $\theta_{IE}('')$ \\
\hline \hline \\
F350LP & 2455205.43 & 0.34 $\pm$ 0.01 & 0.13 $\pm$ 0.02  \\
F350LP &  2455233.52 & 0.32 $\pm$ 0.01 & 0.10 $\pm$ 0.02 \\
F350LP & 2455257.48  & 0.33 $\pm$ 0.01 & 0.11 $\pm$ 0.02 \\
F350LP & 2455296.04 & 0.33 $\pm$ 0.01 & 0.13 $\pm$ 0.02 \\
F555W & 2455205.35 & 0.32 $\pm$ 0.01  & 0.12 $\pm$ 0.02 \\
F555W & 2455233.44 & 0.31 $\pm$ 0.01 & 0.11 $\pm$ 0.02  \\
F555W & 2455244.37 & 0.32 $\pm$ 0.01 & 0.11 $\pm$ 0.02 \\
F555W & 2455257.35 & 0.32 $\pm$ 0.01 & 0.10 $\pm$ 0.01 \\  
F555W & 2455295.96 & 0.32 $\pm$ 0.01 & 0.11 $\pm$ 0.02 \\
F814W & 2455233.57 & 0.35 $\pm$ 0.01  & 0.09 $\pm$ 0.03\\
F814W & 2455263.82 & 0.32 $\pm$ 0.01 & 0.08 $\pm$ 0.03\\
\hline \hline
\end{tabular}
  \end{table*}


To independently confirm the extent of the central echo, the FWHM of the source was determined using the radial profile task ($r$ mode, which uses a radial profile Gaussian fit) in IMEXAMINE in IRAF  \citep{Tody}. This analysis was successful on all echo images but F814W, exhibiting the same trend in each.  The fit did not converge in the F814W filter due to the weaker signal, and only an estimation could be obtained.  The average FWHM of the local stars was 2.63 $\pm$ 0.4 pixels compared to the average inner echo FWHM of 4.03 $\pm$ 0.3 pixels. Thus, the inner echo has a FWHM 1.53 times extended relative to the local stars in the field.  This test illustrates the broadened feature of the source and definitively refutes the notion that the central component is simply a background star. The broadening can be interpreted as nebulosity caused by a light echo.

\subsection{LIGHT ECHO MAGNITUDE}

Light echo magnitudes were determined using the fits mentioned previously and are consistent with the values obtained by estimating by eye the sizes of circular apertures and summing the counts enclosed.

The counts were converted to VEGA magnitudes using WFC3 header keyword PHOTFLAM,  the mean flux density that produces one count per second in observing mode, and the filter zero points provided.\footnote[3]{http://www.stsci.edu/hst/wfc3/phot\_zp\_lbn.} The PHOTFLAM values used in our conversions were 5.297E-20 for F350LP, 1.865E-19 for F555W, and 1.514E19 for F814W (in units of erg cm$^{-2}$\AA$^{-1}$).

\begin{table*}[htbp]
 \caption[Light Echo Magnitudes]{\hspace{135pt}Light Echo Magnitudes\footnotemark[4]}
\label{table:magtot}
\centering
\begin{tabular}{ccccccc}
\hline \hline
\\Filter & JD (days) & Magnitude$_{OE}$ & Magnitude$_{IE}$ & Magnitude$_{Total}$ \\
\hline \hline \\
F350LP & 2455205.43 & 24.477 $\pm$ 0.065 & 25.202 $\pm$ 0.199 & 24.028 \\
F350LP &  2455233.52 & 24.611  $\pm$  0.066 & 25.369 $\pm$ 0.188 & 24.173\\
F350LP & 2455257.48  & 24.552  $\pm$  0.066 & 25.327 $\pm$ 0.189 & 24.119\\
F350LP & 2455296.04 & 24.832 $\pm$ 0.085 & 25.337  $\pm$ 0.217 & 24.303 \\
F555W & 2455205.35 & 24.511 $\pm$  0.057 & 25.109  $\pm$ 0.171 & 24.017 \\
F555W & 2455233.44 & 24.591  $\pm$ 0.072 & 25.204  $\pm$ 0.199 & 24.103 \\
F555W & 2455244.37 & 24.600  $\pm$ 0.064 & 25.223  $\pm$ 0.180 & 24.115 \\
F555W & 2455257.35 & 24.575  $\pm$ 0.063 & 25.259  $\pm$ 0.167 & 24.111 \\  
F555W & 2455295.96 & 24.832  $\pm$ 0.084 & 25.213  $\pm$ 0.182 & 24.253 \\
F814W & 2455233.57 & 25.058  $\pm$ 0.158 & 25.962  $\pm$ 0.374 & 24.666 \\
F814W & 2455263.82 & 24.819  $\pm$ 0.119 & 26.250  $\pm$ 0.452 & 24.562 \\
\hline \hline
\end{tabular}
  \end{table*}
\footnotetext[4]{All magnitudes listed are in VEGA magnitudes}

The magnitudes of the inner and outer echoes listed by filter and date are listed in Table~\ref{table:magtot}.  The total magnitude of the echoes was used in plotting the $HST$ observations on the light curve of SN 2007af (Figure~\ref{fig:LC}) since the ground-based observations would not resolve two separate components.

The magnitude difference between the inner and outer echoes changes with filter.  For the F350LP filter, there is a $\sim$0.6 mag difference between the outer and inner echo values and $\sim$0.7 mag difference in the F555W filter.  However, for the F814W band, this value increases to $\sim$1.0 mag. Even with the lower sensitivity in the $ wide$  $I$ filter, this noticeable change implies a change in dust reflecting properties. This behavior seems inconsistent for two light echoes created in the same manner, which could suggest a different mechanism for producing the inner echo.

\section{PHYSICAL MODEL}
\label{sec:phys}

Light echoes are effective probes of the surrounding material, and simple applications yield ample information about the source. From the geometry of the echoes, a distance between the scattering medium and SN can be inferred, and the color of the echoes reveals information about the scattering effects and dust properties.

\subsection{DUST DISTANCE}

Analytical treatment of the light echo phenomenon (see \citet{Couderc} for more rigorous derivation) states that a dust sheet intersecting the paraboloid of constant time delay creates a circular ring echo centered on the source solely dependent on the time ($t$) since peak light and the distance to the host galaxy ($D$), assuming a dust sheet perpendicular to the line of sight. The distance between the dust sheet and the SN can be simply determined using Equation~\ref{eq:dust}, where $\theta$ is the angular radius of the light echo. 

 \begin{eqnarray}
 d \approx \frac{D^2 \theta^2 - (ct)^2}{2ct}
\label{eq:dust}
\end{eqnarray}

The distance from the dust sheet to the SN for the outer echo was determined using $D$ = 24 Mpc to NGC 5584 \citep{Simon}.  For comparison, the distance estimate based on a joint analysis of the Cepheid and SN data by \citet{Riess}  yields $D$ = 23 $\pm$ 0.7 Mpc after correction for the revised maser distance to NGC 4258 \citep{Hump}.  The dust sheet distances can be found in Table~\ref{table:ds}.  The average distance from the SN to the dust sheet is 786 $\pm$ 63 pc, implying that ISM dust produced the echo.

The inner echo is consistent with both a CSM and ISM sheet and represents the smallest and largest dust distances allowed by the previous analysis. The inner echo dust distance was calculated assuming a circumstellar sheet (d$_{IECSM}$(pc)). Using the proposed solution of backscattered light, a dust sheet located behind the SN exists at a location determined solely by the time delay of the echo. The initial light from the SN traveled outward, scattered off the dust sheet behind the explosion site, and returned back toward the observer. Therefore, half of the time delay determines the distance to the dust sheet.  The average inner echo distance from the SN to the sheet  is 0.45 $\pm$ 0.01 pc in this scenario.  Alternatively, the results for the ISM dust sheet scenario are listed in the last column using the same labeling convention. The inner echo SN-foreground dust sheet distance  results in an average of 89 $\pm$ 24 pc.  
 
\begin{table*}[ht]
 \caption[Dust Sheet Distances]{\hspace{150pt}Dust Sheet Distances}
\label{table:ds}
\centering
\begin{tabular}{ccc  c ccc}
\hline \hline
\\Filter & JD (days) & d$_{OE}$(pc) & d$_{IECSM}$(pc)  & d$_{IEISM}$(pc)\\ 

F350LP & 2455205.43  & 891& 0.43 & 130\\
F555W & 2455205.35 & 790 & 0.43 & 110\\
F555W & 2455233.44 & 721 & 0.44 & 90\\
F350LP & 2455233.52  & 769 & 0.44 &  74\\
F814W & 2455233.57 & 920 & 0.44 & 60\\
F555W & 2455244.37 & 761 & 0.45 & 89\\
F350LP & 2455257.48 & 800 & 0.46 & 88 \\
F555W &  2455257.35 & 752 & 0.46 &  88\\
F814W & 2455263.82  & 748 & 0.46 & 46\\
F350LP & 2455296.04 & 772 & 0.47 & 119\\
F555W & 2455295.96 & 726 & 0.47 &  85\\ 
\hline \hline
\end{tabular}
  \end{table*}

\begin{figure}[h]
\begin{center}
\includegraphics[width=3.5in]{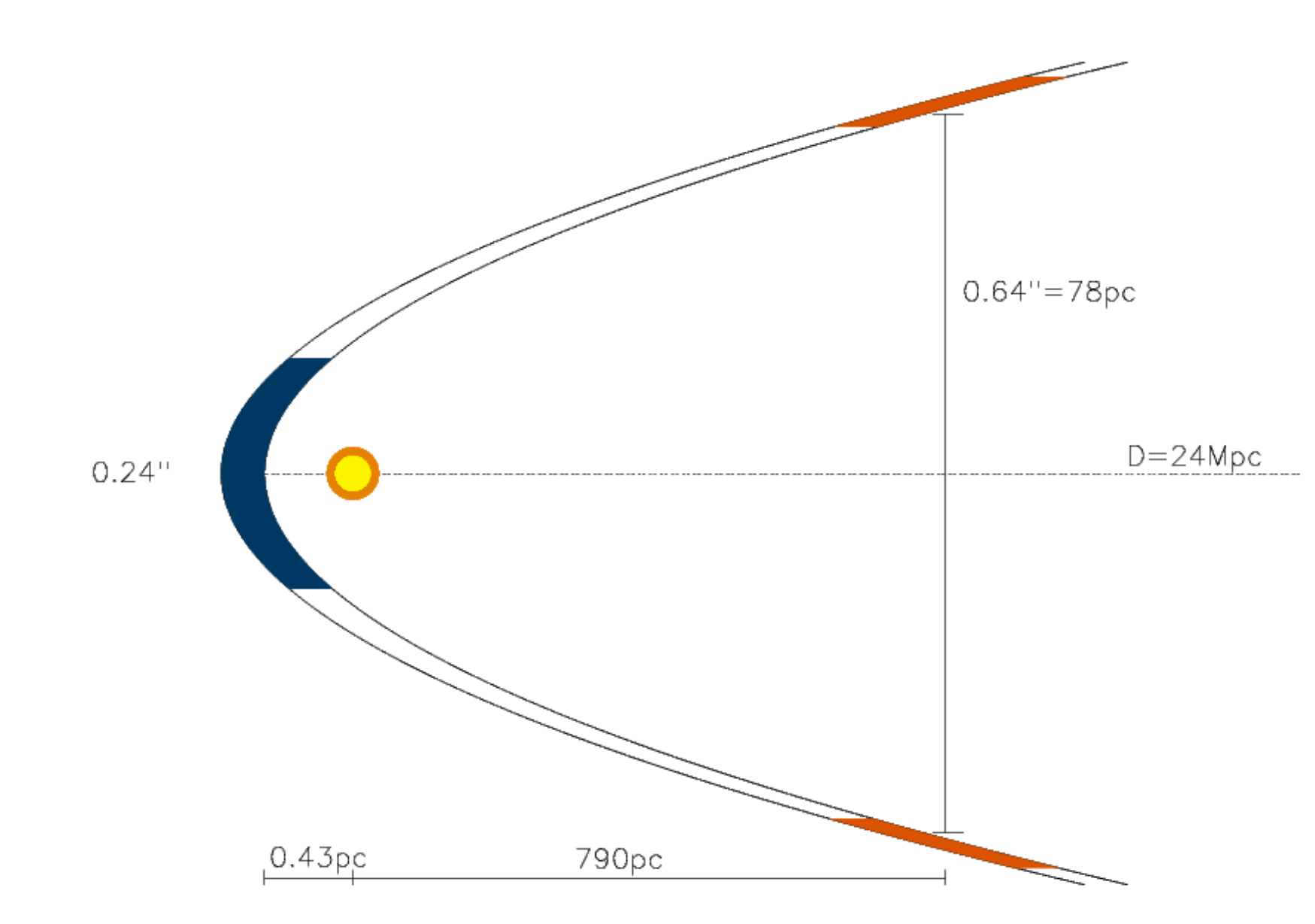}
\end{center}
\caption[Dust Geometry for SN 2007af Echoes-CSM]{The geometry of the dust that produced the inner echo (blue) and outer light echo (orange) of SN 2007af (not to scale) in the backscattering, CSM scenario.  This figure focuses on the January F555W epoch of the light echo, with $\theta_{IE}$ = 0.12$''$, and $\theta_{OE}$ = 0.32$''$.  The figure shows the foreground dust sheet for the outer echo and a secondary dust sheet 0.44 pc behind the SN (shown in yellow).}
\label{fig:dust}
\end{figure}

\begin{figure}[h]
\begin{center}
 \includegraphics[width=3.5in]{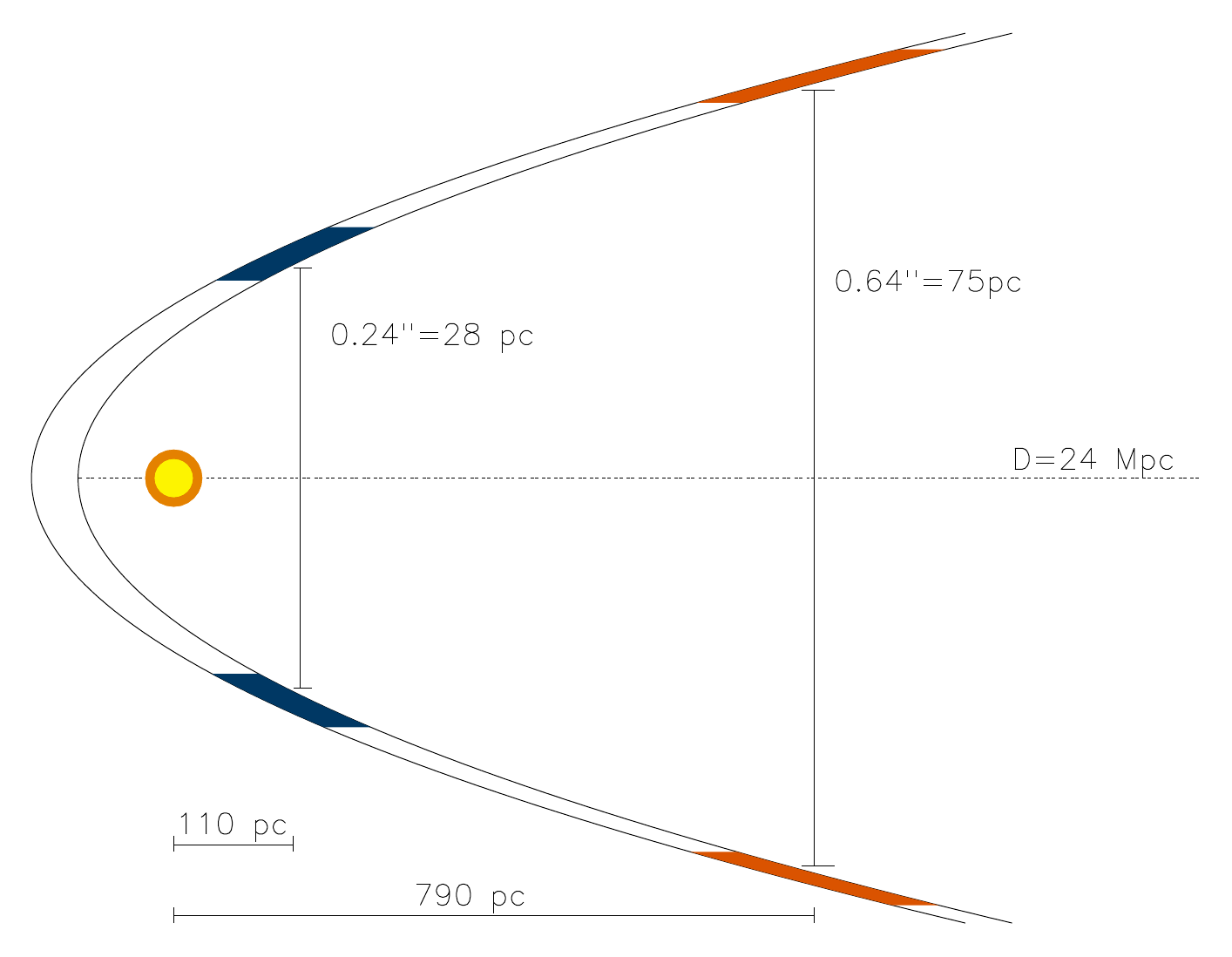}
\end{center}
\caption[Dust Geometry for SN 2007af Echoes-ISM]{The geometry of the dust that produced the inner echo (blue) and outer light echo (orange) of SN 2007af (not to scale) for the forward scattering, ISM scenario.  The outer echo for the January epoch (F555W) was created from a dust sheet 790 pc away from the SN, and the inner echo dust sheet  was located 110 pc in front of the SN (shown in yellow).}
\label{fig:dust2}
\end{figure}

A schematic for the dust geometry (not to scale) for the January epoch in the F555W filter can be seen in Figure~\ref{fig:dust} for a CSM inner echo. Because the distance from the SN to Earth is much greater than the distance between the SN and the dust, we can approximate the geometry of the phenomenon as a paraboloid. The SN is located at the focus of the paraboloid, and the echo rings are created from the intersection between the dust sheet and the paraboloid \citep{Patat05}.  The orange shaded region shows the intersection that results in the outer echo, and the blue shaded region is the intersection causing the inner echo.  Figure~\ref{fig:dust2} shows the same epoch but with an ISM origin for both echoes.

\subsection{DUST COLOR ANALYSIS}

As reported in \citet{Simon}, SN 2007af peaked on 2007 March 14.76 $\pm$ 0.12d (JD = 2454174.26).  Using the \citet{Patat05} model, the initial light is treated as a flash. The signal received by the observer at a later time $t$ is the sum of photons from the SN in the range of time from 0 -- $t$. The global flux measured by the observer is the echo flux combined with the photons that reached the observer on the direct path from the SN, which is then extinction corrected. The light echo flux in a particular passband is given by $F$ = $L$$_{o}$$n$$_{o}$$C$$_{scat}$$\Delta$t$_{SN}$e$^{-\tau_{eff}}$$G$($t$), where $L$$_{o}$ is the SN signal, n$_{o}$ is the number density of scattering particles, and  $C$$_{scat}$ is the scattering cross-section of the dust grains. The flash duration is  $\Delta$t$_{SN}$, $\tau_{eff}$ is the weighted optical depth for the LE at a given time, and $G$($t$) is a wavelength- and time-dependent function related to the dust geometry.

The color of this echo can be predicted using the single scattering approximation model taken from \citet{Patat05}.  The predicted color of the echo (Equation~\ref{eq:color})  was determined by folding the F555W and F814W WFC3 transmission filters with a peak Branch normal SN Ia spectrum scattered with \citet{Draine} dust cross-sections \citep{Template} and weighted by the flash duration in each band.  These are compared to the Vega spectrum folded with the same filter transmission functions.  The observed and theoretical colors are written in terms of F555W -- F814W ($V$ -- $I$).  A normal peak spectrum was used because reddening corrections are then unnecessary, and the peak SN 2007af spectrum did not extend to the blueward wavelengths of the F555W transmission filter.  The flash durations were estimated using the peak light curve of SN 2007af.  We chose the flash duration to be the number of days, centered around maximum, within one magnitude of the peak value.    From Figure~\ref{fig:LC}, the flash durations of $V$ and $I$ were estimated at 0.08 yr (30 days) and 0.11 yr (40 days), respectively.  The $I$ band has a longer flash duration due to the second maximum in the light curve. Longer flash durations would lead to thicker echoes in the $I$ filter.

\begin{equation}
(V - I)  \approx   - 2.5log\frac{F_{V} F^{Vega}_I}{F^{Vega}_{V}F_{I}} 
\label{eq:color}
\end{equation}

Some dust parameters used are listed in Table~\ref{table:params} for illustration, where the third column is the scattering cross section for forward scattering  ($\theta$ = 0$^{\circ}$), and the fourth column values are for backward scattering ($\theta$ = 180$^{\circ}$)\footnote[5]{ftp://ftp.astro.princeton.edu/draine/dust/mix/}.  $C$$_{scat}$ corresponds to the differential scattering cross section per $H$ nucleon, which determines the scattering properties of a particular dust mixture.  The values listed are the ones closest to the central wavelength for 555W and 814W and merely illustrate the relation between forward and backscattering cross section.  For our analysis, we integrated over all wavelengths.

\begin{table*}[htbp]
 \caption[Dust Parameters]{\hspace{155pt}Dust Parameters}
\label{table:params}
\centering
\begin{tabular}{cccccc}
\hline \hline
\multicolumn{2}{c}{} & \multicolumn{1}{c}{$\theta$ = 0$^{\circ}$\footnotemark[6] }  &                      \multicolumn{1}{c}{$\theta$ = 180$^{\circ}$}
 \\Dust Type & Wavelength (\AA) & C$_{scat}$(cm$^{2}$) & C$_{scat}$(cm$^{2}$) \\ 
\hline \hline
MW R$_{V}$ = 3.1 & 5470 & 1.68E-22 & 7.30E-24 \\
MW R$_{V}$ = 3.1 & 8020 & 6.27E-23 & 5.88E-24 \\
LMC R$_{V}$ = 2.6  & 5470 & 5.32E-23 & 1.51E-24 \\
LMC R$_{V}$ = 2.6 & 8020 & 1.96E-23 & 1.10E-24 \\
SMC Bar R$_{V}$ = 2.9 & 5470 & 2.38E-23 & 9.81E-25 \\
SMC Bar R$_{V}$ = 2.9 & 8020 & 8.85E-24 & 6.39E-25 \\
\hline \hline
\end{tabular}
  \end{table*}
\footnotetext[6]{$\theta$ = 0$^{\circ}$ corresponds to forward scattering, and $\theta$ = 180$^{\circ}$ is backscattering. }

\begin{table*}[htbp]
 \caption[Color Values]{\hspace{55pt}Comparison Between Peak, Echo Color, and Model Predictions}
\label{table:color}
\centering
\begin{tabular}{cc|cc|c}
\hline \hline
\multicolumn{1}{c}{} & \multicolumn{1}{c}{Observed} &  \multicolumn{1}{c}{} &   \multicolumn{2}{c}{Theoretical}
\\Epoch &  ($V$ -- $I$) (mag) & Dust Type & $\theta$ = 0$^{\circ}$&  $\theta$ = 180$^{\circ}$ \\ 
\hline \hline
Peak &  -0.327  & MW R$_{V}$ = 3.1 & -1.124 & -0.291\\
Outer Echo & -0.317 $\pm$ 0.104 & LMC R$_{V}$ = 2.6 & -1.122 & -0.561  \\
Inner Echo &  -0.904 $\pm$ 0.304 & SMC Bar R$_{V}$ = 2.9 & -1.127 & -0.545 \\
\hline \hline
\end{tabular}
  \end{table*}

The actual echo colors using the averages measured in each filter was compared to the predicted model colors  (Table~\ref{table:color}). Also listed is the peak dereddened color of SN 2007af (-0.327 mag), which was obtained by integrating the peak SN spectrum over the filter responses.  The results of the \citet{Patat05} model color for forward and backward scattering calculations for the three dust types are listed in the last two columns. All values listed are in the VEGA magnitude system. 

Forward scattering replicates the inner echo color at the far uncertainty limit, but does not discriminate between dust types due to large uncertainties. The outer echo color does not match predictions, and thus, abnormal dust must be considered.  It is interesting to note that the outer echo color has not changed drastically since peak, but that is not the case for the inner echo.   Dust grains are not expected to isotropically scatter light from an echo. Small angle scattering is more efficient.  Thus, forward scattering is the preferred mechanism, which is consistent with the inner echo, but due to our uncertainties, backscattering cannot be ruled out, especially in the case of the outer echo.  We also tried a secondary method of predicting the echo color by binning the light curve and summing over time bins,  with similar results.

Recently, lower R$_{V}$ value have been reported for SNe Ia. Scattered light (which causes echoes) tends to reduce the ratio of total-to-selective extinction in the optical \citep{Wang05}. The ratio is significantly lower at longer wavelengths than 3000\AA. Abnormal dust has been cited for adopting a lower R$_{V}$ value (1.48 $\pm$ 0.06) in the case of SN 2006X \citep{Wang08}. \citet{Goobar} published simulations that yielded values from R$_{V}$ = 1.5 -- 2.5, on the basis that lower values originate from the semi-diffusive propagation of the photons near the location of the SN explosion. Eighty SNe Ia were observed with E(B -- V) $\leq$ 0.7 mag. From this study, an average reddening law of R$_{V}$ = 1.75 $\pm$ 0.27 was derived. Focusing on the 69 SNe with color excess,  E(B -- V) \textless \hspace{1pt} 0.25 mag, produced a total-to-selective ratio value of R$_{V}$ $\sim$ 1, signifying the reddening in SNe Ia may be more complex than previously thought \citep{Nob}. The reddening of SN 2007af fits well under both limits, which could explain the difference between the actual color measured and the predictions using higher R$_{V}$ values.

\subsection{COMPARISONS}

\begin{figure}[htp]
\begin{center}
\includegraphics[width=3.3in]{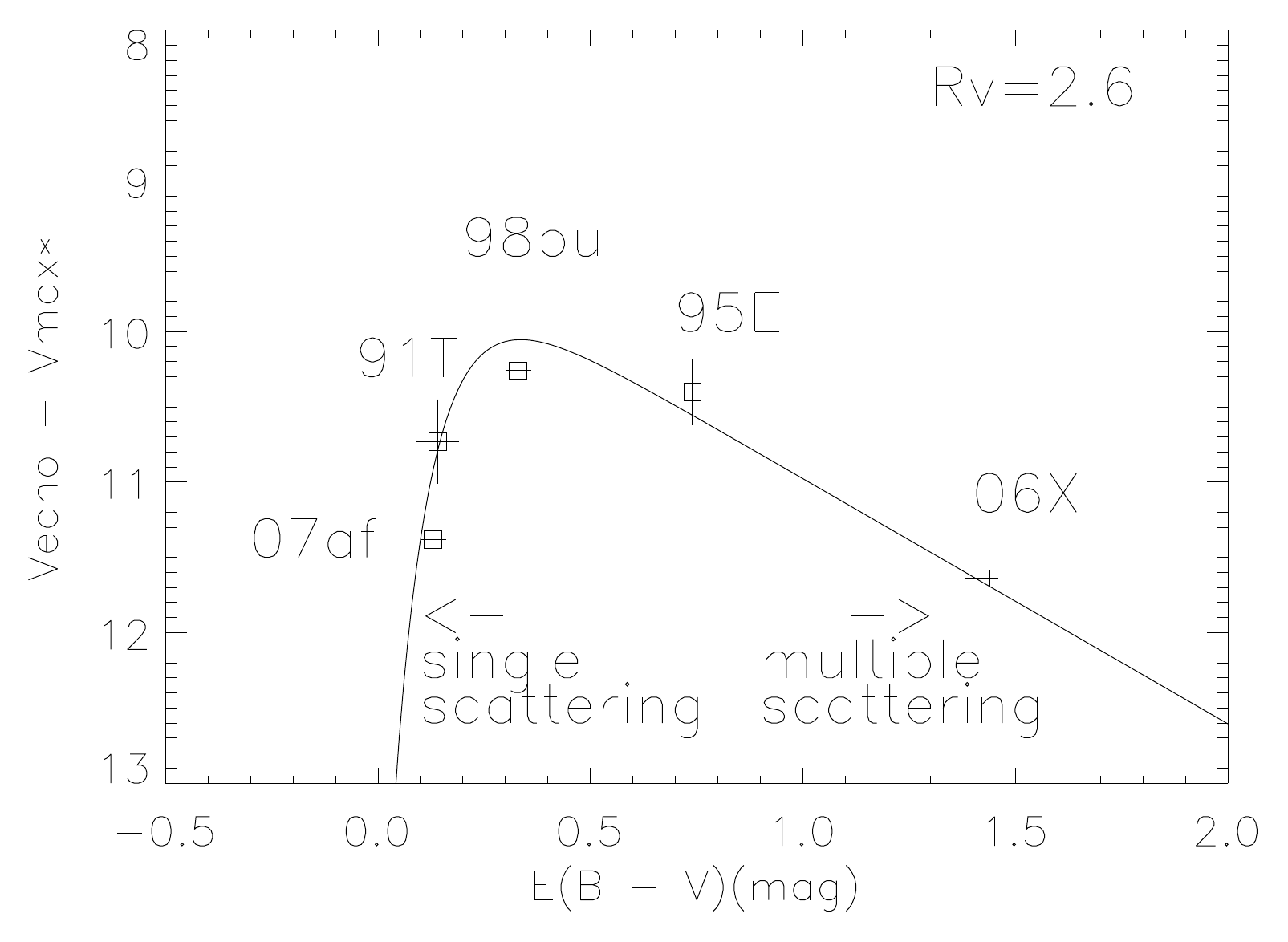}
\end{center}
\caption[Patat Model Compared to SNe Ia Light Echoes]{The difference in extinction corrected $V$ peak to echo magnitude (total magnitude in the case of SN 2007af) compared to the \protect\citet{Patat05} model for single and multiple scattering using a dust law of R$_{V}$ = 2.6.  The figure shows the similarity between all SNe Ia echoes and the excellent agreement to the model.} 
\label{fig:patat2}
\end{figure}

A relation between E(B -- V) and $\Delta$m, the peak-echo magnitude difference accounting for extinction, can be determined assuming the relationship between scattering and absorption is known. Table~\ref{table:snia} lists the values of all reported SNe Ia light echoes including SN 2007af (using R$_{V}$ = 2.6). A lower R$_{V}$ was adopted after the dust analysis results. The combined magnitude for SN 2007af converted to standard Bessell magnitudes was used for this analysis \citep{Bessell}. The first three rows were taken from \citet{QuinnGarn}, and SN 2006X values were reported in \citet{Wang08}.  E(B -- V) for SN 2007af comes from \citet{Simon}. The values are plotted in Figure~\ref{fig:patat2}, showing the excellent agreement between the SNe and the \citet{Patat05} model for single and multiple scattering. For low optical depth values ($\tau$ $\leq \hspace{1pt} $1.0), the light echo phenomenon is well described by single scattering. Auto-absorption and attenuation becomes significant for $\tau$ $\textgreater$ 1.0, and multiple scattering must be considered.

\begin{table*}[ht]
 \caption[SN Ia Light Echo Comparisons]{\hspace{130pt}SN Ia Light Echo Comparisons}
\label{table:snia}
\centering
\begin{tabular}{cccccccc}
\hline \hline
\\Supernova & E(B -- V)(mag) & $\Delta$V(mag)\\
\hline \hline
SN 1991T & 0.14 $\pm$ 0.05 & 10.73 $\pm$ 0.28\\
SN 1998bu & 0.33 $\pm$ 0.03 & 10.26 $\pm$ 0.22\\
SN 1995E & 0.74 $\pm$ 0.03 & 10.40 $\pm$ 0.22\\
SN 2006X & 1.42 $\pm$ 0.04 & 11.64 $\pm$ 0.07\\
SN 2007af & 0.13 $\pm$ 0.03 & 11.38 $\pm$ 0.13\\
\hline \hline
\end{tabular}
  \end{table*}

\citet{Patat05} proposed that LE luminosity is inversely proportional to SN-dust distance ($d$) and directly proportional to the optical depth. The peak-to-echo magnitude relationship in the single scattering approximation is shown in Equation~\ref{eq:patat}.  The multiple scattering approximation to $\Delta$m is given in Equation~\ref{eq:patat2} and follows an exponential decay. These relationships hold for $d$ measured in ly.

 \begin{eqnarray}
  \hspace{40pt} \Delta m \approx -2.5log(0.3\frac{{ \tau}}{d}) \hspace{20pt}  (0 \hspace{2pt} \textless \tau \leq 1)
\label{eq:patat}
\end{eqnarray}

 \begin{eqnarray}
 \hspace{40pt}  \Delta m \approx -2.5log(\frac{0.3}{d}e^{-\tau}) \hspace{28pt}                (\tau \hspace{2pt} \textgreater 1)
\label{eq:patat2}
\end{eqnarray}

Peak magnitude extinction correction was performed using ratios of total-to-selective extinction of R$_{V}$ = 2.6 $\pm$ 0.1 and R$_{I}$ = 1.48 $\pm$ 0.1. The inner echoes have not been extinction corrected because the dust associated with the outer echo would be the dominant cause of the extinction of peak light. The optical depth was calculated using the ISM distances for the inner echo and outer echo. The $V$ optical depths are 0.15 and 0.01 for the outer and inner echoes, respectively.  For $I$, the values are 0.16 and 0.01.  From the large dispersion in the values, it is clear that the dust sheet for the inner echo must be optically thin.  This confirms the outer echo dust sheet dominance to the line-of-sight extinction. Because of the low optical depth, SN 2007af fits within the single scattering approximation.

SN 2007af featured a lower E(B -- V) at peak than average SN Ia and the lowest value of all light echo candidates to date. The fact that a significant portion of the sample of SNe Ia found with light echoes have low extinction values is surprising. SN 2007af showed nothing in its peak behavior to suggest the possibility of an echo at late epochs. The production of a light echo from such environments needs to be further explored. 

The comparison of SN 2007af with other SNe Ia light echoes is challenging due to the multiple components. SN 1998bu is the only other clear case of a dual-feature light echo, and the inner echo was also proposed to have been caused by circumstellar dust \citep{Garn}. Unfortunately, that echo and all the other SNe Ia candidates were not captured with the $I$ filter, so a rigorous comparison cannot be made. SN 2006X was the most comparable case, since it was observed in the F775W filter with $HST$. The echo from SN 2006X showed a distinct wavelength dependence, with more light scattered from shorter wavelengths. The team credits this trend to small dust size \citep{Wang08}  This pattern is also observed in SN 2007af and can be explained with shorter wavelengths being more efficient scatterers. Since the time delay and distance to the SNe are widely different for all of the SNe cases, we are unable to make comparisons other than the ones previously stated.

\section{DISCUSSION}
\label{sec:disc}

The extended profile of the inner source, brightness, and plateau in luminosity suggest a secondary echo. A broadened FWHM does not definitively prove the existence of an inner echo, but it does eliminate the possibility of a star. However, even the bluest of stars would not replicate the $V$ -- $I$ color observed \citep{Ducati}. A galaxy directly behind the SN could still explain the excess emission.  However, irregular galaxies are typically the bluest of all galaxies with $B$ -- $V$$\sim$0.4, which is not consistent with our analysis \citep{Hunter}. The FWHM  analysis with local stars was performed by \citet{Wang08} on SN 2006X, and the broadening found from that echo was only 1.10 times the local stars. Our inner echo was considerably more extended. No source (echo or other) is evident in the F160W image, which is expected for a light echo since scattering is not efficient in the NIR.  Also, ground-based photometry using pre-explosion subtraction images still show an obvious flattening, making a background galaxy scenario unlikely.  

The limiting magnitude of the F160W filter from the WFC3 Instrumental Handbook for a one hour exposure is 26.6 mag \citep{Dress}.  Thus, an object must be fainter than this value in IR to explain the phenomenon. 

The light echo color for the inner echo is roughly consistent with forward-scattered light off ISM dust, but other scattering angles cannot be eliminated. From our results, we cannot determine whether the inner echo is ISM or CSM in origin, so we will outline the arguments from both sides. 

At early epochs, the Na D absorption showed no variation, providing no supporting evidence of CSM material.  If the inner echo is of circumstellar origin with the scattering material located behind the SN,  the same line variance as seen in SN 2006X would not be observed. \citet{Blondin} reported the second case of variable Na I lines from SN 1999cl.  SN 1999cl and SN 2006X were the two most reddened objects in the sample, suggesting the possible relation between high amounts of reddening and time-variable lines. Perhaps the low-reddening of SN 2007af produced a weaker Na D line, which would be more difficult to detect or, as the paper suggests, there exists a preferred line-of-sight to observe the spectral line variability.

 The upper limit distance derived between the SN and the dust is $\sim$90 pc, which suggests an ISM origin of the inner echo.   SN 2006X and SN 1998bu have light echoes argued to come from the circumstellar environment, but both have distances to the dust well inside our value (\textless \hspace{1pt}10 pc for SN 1998bu \citep{Capp, Garn}).  Further investigation is necessary to explain how the inner and outer echoes created from ISM dust sheets could produce different colors. 

Abnormal dust should be considered in the case of these echoes.  \citet{Goobar} argues that CSM dust around SNe Ia lowers R$_{V}$ values due to multiple scattering of photons, which could explain the color difference between inner and outer echo.

A CSM echo suggests the single-degenerate nature of the progenitor system. However, \citet{Patat06} argue that the existence of circumstellar material around SNe Ia requires the dust to be optically thick, which is not what was found in our analysis. The reason is the dust near the SN might be destroyed by the original SN flux, making a circumstellar echo unlikely.  However, the CSM can be arbitrarily thick if backscattered, and therefore, not on the line of sight.

\section{CONCLUSIONS}
\label{sec:conc}

We present the two-component light echo discovery from SN 2007af in NGC 5584 detected in sequenced images from \textit{HST} three years after explosion. Out of the four filters utilized in the Cepheid campaign, three show the ring-like structure of an outer echo and extended central inner echo component. Ground-based observations taken at the same epoch also show the same uncharacteristic late  brightness of the normal SN Ia that remains constant, which is attributed to the presence of a light echo. Our observations in the nebular phase, where the light from the SN has faded sufficiently enough to let the light echo dominate the emission, are vital for probing the SN environment.

 Using the difference between the extinction-corrected peak magnitude and echo magnitude, the optical depth of the dust was estimated, showing the optically thin nature of both dust sheets. The total magnitude difference ($\sim$11.3 mag in $V$) was consistent with reddening values inferred from peak. By using the \citet{Patat05} model, the collection of SNe Ia light echoes was well-approximated. SN 2007af fit the single scattering approximation of this model.

The outer echo F350LP magnitude was found to be $\sim$24.6 mag, $\sim$24.6 mag in F555W, and the F814W magnitude was $\sim$25.0 mag. The distance from the SN to the dust sheet that created the outer echo was 786 $\pm$ 63 pc. The dust color was not consistent with Galactic dust in the forward-scattering scenario. The color was better replicated by back-scattering, but the inefficiency in that scattering method suggests abnormal dust.  The brighter outer echo compared to the inner echo suggests the outer echo dust sheet dominates the extinction in the SN peak light curves.  

The inner echo of SN 2007af is the best imaged candidate for a CSM echo. Even though the \citet{Simon} publication saw no significant evidence of circumstellar material using high- and low- resolution spectroscopy, they do note that there could be a preferred geometry for observing the time-variance in spectral lines. The FWHM of the inner echo was 1.5 times broader than the field stars, illustrating the extended nature of the central component and confirming the source was not a star at the location of the SN. No galactic source was found at the location of the IE in the F160W image, supporting the argument of a secondary echo. The inner echo  magnitude in the F350LP filter was $\sim$25.3 mag, $\sim$25.2 mag in F555W, and $\sim$26.1 mag in F814W. The SN to dust distance was estimated to be 89 $\pm$ 24 pc for a dust sheet located in front of the SN.  For a dust sheet located directly behind the SN, a distance of 0.45 $\pm$ 0.01 pc was determined. Although the inner echo color was consistent with forward scattering, backscattering could not eliminated, and the color was quite different from the outer echo, suggesting that even lower total-to-selective extinction ratios, other geometries, or atypical dust must be considered.

  \bibliography{07af}

\end{document}